# The Analysis of a Proposed Experiment to Measure the Speed of Gravity in Short Distances

[1,2]Carlos Frajuca, [2]Fabio da Silva Bortoli and [3]Nadja Simão Magalhaes

[1] IMEF, Federal University of Rio Grande, Rio Grande 96203-900, Brazil.

[2] Campus Sao Paulo, Sao Paulo Federal Institute, Sao Paulo 01109-010, Brazil.

[3] Department of Physics, Federal University of Sao Paulo, Diadema 09913-030, Brazil.

**Abstract**

In order to investigate the speed of gravitational signals travelling in air or through a different medium two experiments were designed. One of the experiments contains 2 masses rotating at very high speed and in the other experiment a sapphire bar will vibrate, in both cases they will emit a periodic tidal gravitational signal and one sapphire device that behaves as a detector, which are suspended in vacuum and cooled down to 4.2 K will act as a detector. The vibrational amplitude of the sapphire detector device is measured by an microwave signal with ultralow phase-noise that uses resonance in the whispering gallery modes inside the detector device. Sapphire has a quite high mechanical Q and electrical Q which implies a very narrow detection band thus reducing the detection sensitivity. A new detector shape for the detector device is presented in this work, yielding a detection band of about half of the device vibrational frequency. With the aid of a Finite Element Program the normal mode frequencies of the detector can be calculated with high precision. The results show a similar expected sensitivity between the two experimental setup, but the experiment with the vibration masses is more stable in frequency then it is chosen for the experimental setup to measure the speed of gravity in short distances. Then a more precise analysis is made with this experiment reaching a signal-noise ratio of 10 at a frequency of 5000 Hz.

# Introduction

The speed of gravity in classical theories of gravitation, is the speed that changes in a gravitational field propagate. A change in the distribution of momentum and energy results in subsequent change, at a certain distance, of the gravitational field which it produces. In the theory of general relativity, the "speed of gravity" can be referred to as the speed of a gravitational wave observed by the GW170817 neutron star merger, and is the same speed [1] as the speed of light ($c$). It could provide a final demonstration of the gravity, but as shown below, this is not the case.

Newtonian gravity requires that each particle with mass respond instantaneously to every other particle with mass independently of the distance between them or, according to which, when the mass distribution changes, its gravitational field instantaneously adjusts to the new distribution, making the speed of gravity, in this theory, infinite. Only in the 19th century an anomaly in astronomical observations that could not be explained with the Newtonian gravitational was discovered: the French astronomer Urbain Le Vernier determined in 1859 that the precession of the elliptical orbit of Mercury changes at a significantly different rate then the one predicted by Newtonian gravity [2].

Laplace tried to include a finite speed within Newton's theory in 1805. Based on Newton's law of force he considered a model in which the gravitational field is defined as a radiation field or fluid [3]. The movements of the celestial bodies should be modified in the order $v/c$, where $v$ is the relative speed between the bodies and $c$ is the speed of gravity. Then Laplace calculated that the speed of gravitational interactions is at least $7 \times 10^6$ times the speed of light. This finite speed also leads to some sort of aberration and therefore makes the orbits of the planets unstable.

At the end of the 19th century, many scientists tried to combine the laws of electromagnetism with Newton's law of force. Those theories contain additional terms that maintain the stability of the planetary system. Those models also were used to explain the advance of Mercury perihelion, but without success.

In 1900, Hendrik Lorentz using ether theory and Maxwell equations tried to explain gravity. The result is exactly what is known as universal gravitation, in which the speed of gravity is equal to the speed of light. However, Lorentz showed that his theory is not a problem pointed out by Laplace, because in Lorentz equations only effects in the order $v^2/c^2$ arise. But Lorentz calculation for the value of the Mercury perihelion change was much too low [4].

Henri Poincaré, in 1908 examined the Lorentz gravitational theory and classified it as compatible with the relativity principle, but he pointed out the inaccuracy of the perihelion advance of Mercury [5-7].

Similar models were proposed by Hermann Minkowski and Arnold Sommerfeld. However, these models were eclipsed by Einstein's theory of general relativity [8] which predicts that gravitational radiation should propagate at lightspeed.

The speed of gravity can be obtained from the observation of the binary pulsars PSR 1913+16 and PSR B1534+12 orbital decay rate. The orbits of these binary pulsars are decaying due emission of gravitational radiation. The rate of this energy loss can be measured, and it depends on the speed of gravity, and calculations show that the speed of gravity is equal to the speed of light to within 1% [9]. But, there are two main limitations of the post-Newtonian approximation for describing gravitational wave emission and the motion of binary pulsars: 1) Near the pulsars the gravitational field is strong and the weak-field assumption no longer holds. 2) When gravitational waves are generated (of wavelength λGW) and their back-reaction on the orbit (of size r and period $P_b$), the post-Newtonian approximation is valid only in the close zone (r << λGW = $cP_b$/2), and fails in the radiation zone (r > λGW) where gravitational waves propagate [10,11].

In September 2002,there was an an announce that the speed of gravity was measured indirectly, using data from VLBI measurement of the retarded position of Jupier on its orbit during Jupiter's transit across the line-of-sight of the bright quasar QSO J0842+1835. The authors calculated that the speed of gravity is between 0.8 and 1.2 times the speed of light [12]. Many physicists didn't agree with these claims. For example, some scientists theorise that the experiment was essentially a measurement of the speed of light [13] or the effects were too small to be measured [14].

The detection of GW170817 in 2017, the neutron star inspiral observed through gravitational waves and gamma rays, currently indicates by far the best limit on the difference between the speed of light and that of gravity. Photons were detected 1.7 seconds after the peak of the gravitational wave maximum; the difference between the speeds of gravitational and electromagnetic waves is constrained to between $-3\times10^{-15}$ and $+7\times10^{-16}$ times the speed of light [15]. This result could exclude some alternative theories to general relativity, including variants of the scalar-tensor theory [16,17], instances of Homdescki's theory [18] and Horava-Lifshitz theory of gravity [19-21]. Nevertheless, this result is under some debate as gravitational waves didn't trigger the search for coincidence with other experiments and it is the only detection of this kind although many gravitational wave detections have been recorded.

Besides General Relativity Theory, there is the string theory that provides a speed to gravity and in many versions of string theory the gravity velocity is higher than the speed of light. It will be much more reliable to measure the speed of gravity in an experiment, where the signal could be produced and the effects measured as desired. That is what the authors propose here. Clearly the experiment does not propose to produce gravitational waves as such signals produced in the laboratory are too small to be detected. But a tidal gravitational signal could provide a feasible way to measure the speed of gravity.

The authors are part of the Graviton Group, which is a research group in Brazil devoted to the study of gravity, as part of these studies Gravitational Waves (GW) is the central focus of research. The expertise gained in the field of GW detection projecting the experiment entitled the group with knowledge to design an experiment to measure the speed of gravity. The references [22-39] show the expertise of the authors in microwave electronics, electromechanical transducers, data analysis, high speed machines, vibrational analysis, cryogenics and gravitation.

The knowledgment developed with GW gave the group the expertise and the eagerness to understand gravity. Keeping that in mind the group is developing an experiment to measure the speed of gravity and to do it in short distances. Making the experiment in a short distance allows the possibility of influence of some medium located between the source and the detector changes the results of the experiment.

To reach such a detection two experiments were designed: a quadrupolar distribution of masses will rotate at very high and very stable in time speed or a quadrupolar distribution of masses will vibrate in high frequencies. In the next section the experiments will be described and their limits and sensitivities of the signal will be shown. Then one of the experiments will be chosen for a more complete analysis and the operational frequency will be determined.

## 2. The development of the experiments

The first experimental setup proposed is made of two rotating masses with mass called M, rotation in a radius of value "a´" that is called the emitter and a sapphire bar which is called the detector and is modelled as two masses with mass "m" connected by a spring as can be seen in Fig. 1. A description of how the emitter will look like in reality can be seen in Fig. 2. The detector device is a sapphire bar suspended by its centre in vacuum and cooled down to a temperature of 4.2 K in a liquid helium cryostat. More details can be seen in ref. [40]. The calculated displacement signal in the detector of length b is given by:

$$\Delta b = \frac{24 Q G M_{eff} a^2 b}{\omega^2 r^5} \tag{1}$$

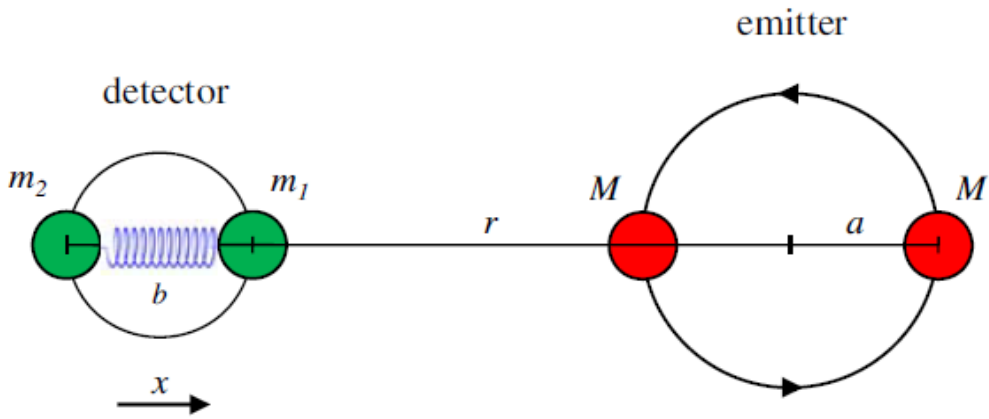

Figure 1. Simplified modelling of the periodic gravitational signal emitting device and the gravitational signal detector. Source: from the authors.

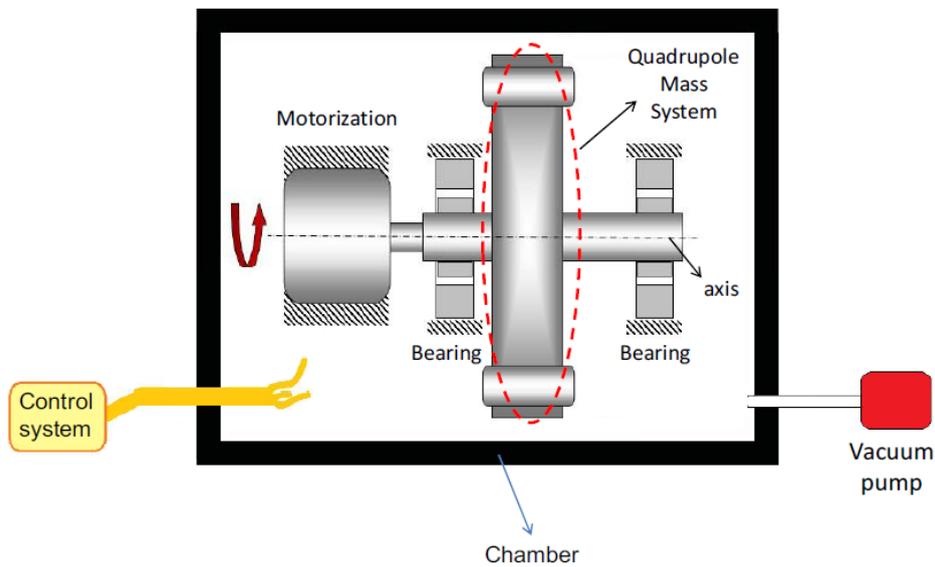

Figure 2. Mechanical system to generate the rotation of the emitter. Source: from the authors.

The second possible experiment was proposed by Fernandes, Gennari and Frajuca [41] and a schematic can be seen in Fig. 3 where the sapphire bars located in the sides have their vibration modes driven by piezoelectric devices (PZT), as the sapphire bars vibrate they emitte a gravitation periodical tidal signal. This signal will excite the vibration modes of the central sapphire bar and its vibration amplitude will be measured by microwave signal pumped with very low vibration phase noise. All the sapphire bars are modelled as a bar-spring system (Fig. 4) and are located in an environment in vacuum cooled to a temperature of 4.2 K. Bodies of mass 'm' (that will be called '$M_{eff}$') are vibrating with a specific amplitude 'a' at a distance 'X' from the detector masses. This scheme is used to calculate the forces between the emitter and the detector. The detector is modelled by two masses 'm' connected by a spring. In ref. [42] appeared the idea to make the experiment in short distances and include the speed of gravity in different mediums. The calculated displacement signal in the sapphire detector is:

$$\Delta b = \frac{24 Q G M_{eff} a b^2}{\omega^2 X^5} \qquad (2)$$

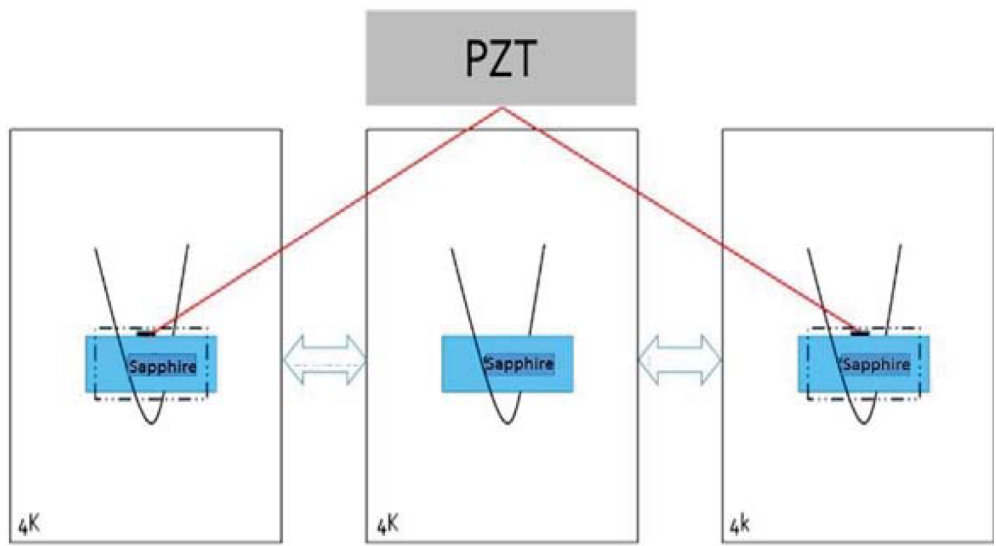

Figure 3: Simplified model of experiment of periodic tidal gravitational signals. The emitter also can be used as a calibrator for GW detectors. Source - From the authors [41].

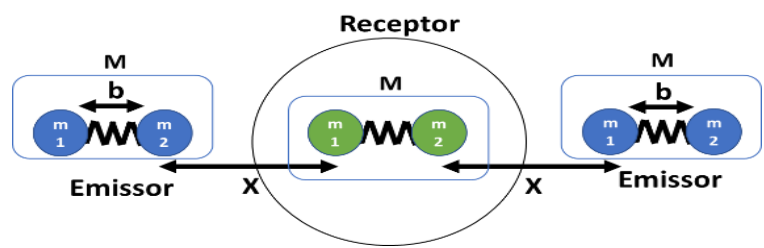

Figure 4. Model of the detector and the emitter of periodic tidal gravitational signals. Source: The authors [41].

As can be seen in the references [41,42] one of the limitations in the measurements is the equipment sensitivity limit, representing the level that the microwaves can measure the vibration of the sapphire bar, it will depend on sapphire bar detection bandwidth, as the sapphire has a very high mechanical Q this bandwidth is intrinsically low. In order to improve this a different shape for the detector is proposed in [43] and can be seen in Fig. 5. This shape works as a three mode detection system, the modes can be seen in Fig. 6, the exact values are not important as it depends in the final design of the experiments, but the detection bandwidth is of the order of one third of the vibration detection frequency, which improves the equipment sensitivity by a great factor. The modes of such device can be seen in [43].

One important point is that the generated signal couples more with the mode with the lower frequency, in this mode all bars are excited by the emissor as they are in phase. In the rest of the work only the forces on the central bar will be considered, the forces on the side bars make the signal stronger.

The microwave electronics to be used in the experiment can be seen in Fig. 7. There is a microwave low phase noise source that goes close to the detection system, is pumped inside the sapphire detector, interacts with the gallery whispering modes, comes out of it, then passes through a microwave suppression system. Then it is amplified and goes to a mixer to restore the vibration frequency of the detection system (decoupled from the original microwave frequency), it is filtered and processed. The phase of the detector vibration is compared to the signal generated by the emitter, this time is compared to the distance between emitter and detector and the gravity speed is calculated.

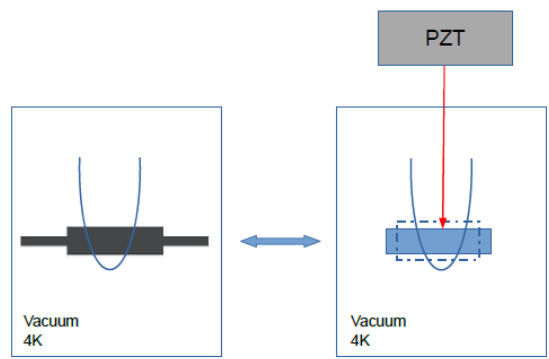

Figure 5. The suspension of the broadband detector system. Source: From the authors [43].

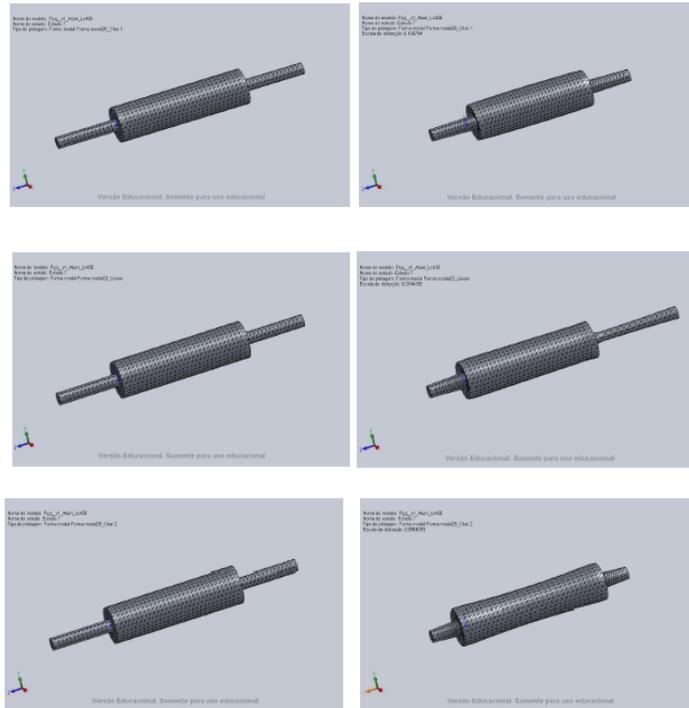

Figure 6. Vibration modes of the broadband detector. The mode on the top all three bars vibrate at the same phase, frequency: 4772 Hz; the mode in the middle the central bar does not vibrate but it oscillates and the side bars vibrate in counter phase, frequency: 5961 Hz; the mode on the bottom the side bars vibrate in counter phase to the central bar, frequency: 7170 Hz. Central bar length is equal to 0.63 m and side bar length is equal to 0.3 m. Source: from authors [43]

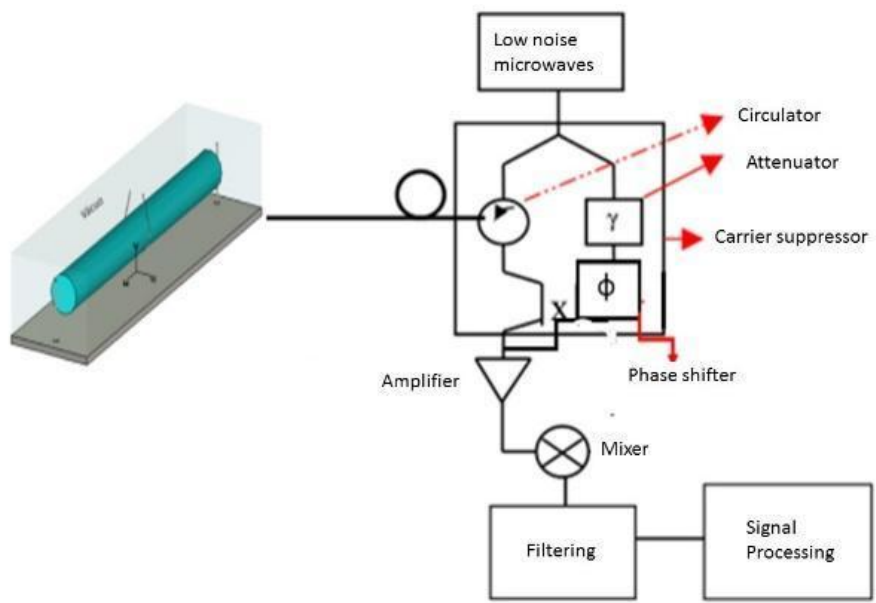

Figure 7. The experiments microwave electronics. Source: from the authors.

# 3. Choosing the experiment mountings

Next phase was to optimise the operating frequency of the experiments (vibrational or rotational) it was done respectively, in [44,45], using that the frequency is given by the size of the detector for the experiment of vibrational masses and the frequency is limited for centrifugal resistance for experiment of rotating masses. In this chapter the electronic series noise and the back action noise were not taken into account as the authors know it will be close to the quantum limit for these characteristics. This will be clear in the next chapter. The both experiment have their sensitivities and limits calculated using the following parameter:

$S_\varphi$ = -160 dBc/Hz at 1 kHz (Microwave phase noise);

$M_{eff}$ = 1 kg (Effective mass of the detector);

$G = 6.674184 \times 10^{-11}$ m$^3$ kg$^{-1}$ s$^{-2}$ (Newton constant);

a' = 0.1 m (Rotation radius for the first experiment)

$a = 10^{-4}$ m (Vibration amplitude of the bars);

b = 0.2 m (Equivalent size of the detectors);

r or X = 1.0 m (Distance between detector and emitter);

BW = 1000 Hz (Adopted frequency bandwidth);

$h = 6.62607004 \times 10^{-34}$ m$^2$ kg s$^{-1}$ (Planck constant);

$f = 10^3$ Hz (Vibrational frequency);

$K = 1.38064852 \times 10^{-23}$ m$^2$ kg s$^{-2}$ K$^{-1}$ (Boltzman constant);

df/dx = 2 x $10^{12}$ Hz/m (Frequency sensitivity of sapphire bar).

As the experiment is connected to vibration, sources of vibration should be taken into account. One of these sources is the thermal noise caused by thermal vibration, when the thermal bath in the detector makes it vibrate coherently. The signal amplitude should be bigger than this noise. The quantum limit shows the amplitude when there is only one phonon in the lattice of the detector, operating above this limit is desirable, otherwise other effects should be taken into account.

For the experiment with vibrating masses, using a operational rotating frequency of 1 kHz, the values for the limits and signal (Δb) are [45]:

Quantum limit: $\Delta b_{QL} = 4 \times 10^{-19}$ m;

Equipment sensitivity limit: $\Delta b_{ES} = 1.6 \times 10^{-18}$ m;

Thermal noise limit: $\Delta b_{th} = 2 \times 10^{-20}$ m;

The signal amplitude is $= 4 \times 10^{-12}$ m.

For the experiment of rotating masses, using the same operational vibration frequency of 1 kHz, the limits and amplitude ($\Delta b$) are the following [44]:

Quantum limit: $\Delta b_{QL} = 2 \times 10^{-19}$ m;

Equipment sensitivity limit: $\Delta b_{ES} = 1.6 \times 10^{-18}$ m;

Thermal noise limit: $\Delta b_{th} = 3.2 \times 10^{-20}$ m;

The signal amplitude is $= 4 \times 10^{-12}$ m.

The expected amplitude value to be measured is value of amplitude to the measured is expected to be [44,45] of the order of:

$$\Delta b = 4 \times 10^{-12} \text{ m.}$$

The both experiments presented a similar sensitivity, as the experiment with the vibrational masses is simpler and safer; it was chosen as the prefered mounting for the experiment. Although this first analysis does not take into account electronic series noise and back action noise, it does not invalidate the choice as those are characteristic only for the detector device.

## 4. The complete analyses of the experiment sensitivity

Chosen the experiment mounting, a complete analysis of the noises and limits is now presented. For this new characteristics are now needed.

$S_{am} = -180$ dBc/Hz at 1 kHz (Microwave amplitude noise);

$P_{inc} =$ (Incident microwave power, to be determined);

$F_{pump} = 10^{10}$ (Microwave signal pump frequency);

$T_{amp} = 10$ K (Effective amplifier temperature).

The sources of noise in this part includes: the microwave phase-noise, the microwave amplitude noise (this is vibration on the sapphire device due to changes in the microwave amplitude, this is the called back-action noise) and the electronics series noise (this represents the level of microwave signal that can be discerned in the amplifier).

For the effective mass of the detector device, a mass of one kilogram was chosen. The dimensions of the detector device depends on these dimensions but for simplicity the diameters of the device changes to maintain its effective mass.

In this part all formulas present the displacements squared to make the comparison of the different sources of noises and limits compatible.

First let's take into account the quantum limit, it's the minimal length that can be measured taking into consideration the uncertainty principle. It´s calculated making the he energy of an harmonic oscillator equal to the energy of one phonon:

$$\Delta b^2_{QL} = \frac{2\hbar}{\omega M_{eff}} \tag{3}$$

Now let's consider the limit imposed by the equipment sensitivity, the signal comes from the sidebands of a microwave signal that lives the central saffire bar in the detector device that acts as a microwave cavity, this kind of transducer is the same one designed to work in gravitational wave detectors and presents the following dependency with detector device frequency [46] in its squared displacement:

$$\Delta b^2_{ES} = (\frac{df}{dx})^{-2} \frac{\omega^4}{\Delta\omega} S\varphi \tag{4}$$

Now it is important to consider the averaged square thermal displacement of the central sapphire bar of the detector device [47].

$$\Delta b^2_{ThN} S_{TH} = \frac{KT}{2\omega M_{eff} Q (\Delta\omega)} \tag{5}$$

Let's now include the back-action noise ($\Delta b_{BAN}^2$) due to variation in amplitude of the microwave pumped signal and the electronics serial noise ($\Delta b_{ESN}^2$) due to noise in the amplifier [46].

$$\Delta b_{BAN}^2 = (\frac{P_{inc}}{\omega^2 M_{eff} 2\pi} \frac{Q_e \frac{df}{dx}}{F_{pump}^2})^2 \frac{\omega^2}{\Delta\omega} S_{am} \quad (6)$$

$$\Delta b_{ESN}^2 = \frac{KT_{amp}}{P_{inc}} \frac{\omega^2}{\Delta\omega} (\frac{F_{pump}}{Q_e \frac{df}{dx}})^2 \quad (7)$$

Comparing the back action noise to the series electronic noise, the best incident power can be found making when the two noises have the same amplitudes and it has the following expression in function of the angular velocity:

$$P_{inc} = 6 \times 10^{-6} \, \omega^{4/3} \quad (8)$$

using this expression for the incident power and using the adopted value for the other characteristics, an expression can be found for $\Delta b_{BA}^2$ and $\Delta b_{ES}^2$.

And finally the the square of the amplitude can be expressed by the following formula:

$$\Delta b_{Signal}^2 = (\frac{24ab^2}{\omega^2} \frac{QGM_{eff}}{x^5})^2 \quad (9)$$

Assuming that has a relationship with b, the central bar of length of 0.63 m has a first frequency of 4772 Hz, giving a ratio of $4.4 \times 10^{-10}$ for $\left(\frac{b}{\omega}\right)^2$. Substituting all the values in the equation, the constant value of $3.6 \times 10^{-27}$ is found. Using this value and the other proposed characteristics a final dependency can be identified and together with the other noise and limits can be plotted together in a graphic. The graphic can be seen in Fig. 8.

Fig. 8 shows that a maximum frequency that the experiment should work is around 5 kHz, in this frequency the signal to noise ratio is about 10 (in metres). Higher frequencies are good because it make easier to build the devices as they can be made smaller and, also, makes easier to differentiate the phase in the emitter from the phase in the detector

The Newtonian noise was analysed as it should not be a problem at the operational frequency of 5kHz [48], or the experiment could be run underground.

The seismic noise was not considered as it can be minimised by making a suspension that isolates the seismic noise in the correct factor [49].

To avoid charge to be built in the experiment, the devices can be submitted to ultraviolet light.

## 5. Conclusions

The experiment with vibrating masses reunites more favourable characteristics as it has a completely stable operating frequency because it only depends on the length of the sapphire device. This experiment is the one chosen to measure the speed of gravity.

The work shows the possibility to measure the speed of gravity in short distances with a signal to noise ratio of about 10 operating at a frequency of around 5 kHz.

The introduction of the broadband detector in the experiment was a great breakthrough, it makes possíble the equipment sensitivity limit close to the quantum and thermal limits of the experiment, otherwise the narrow band of the detector would make the measurements impossible.

Next step is to put together an executive design to mount the experiment and, in the future, use it to discriminate theories of gravity.

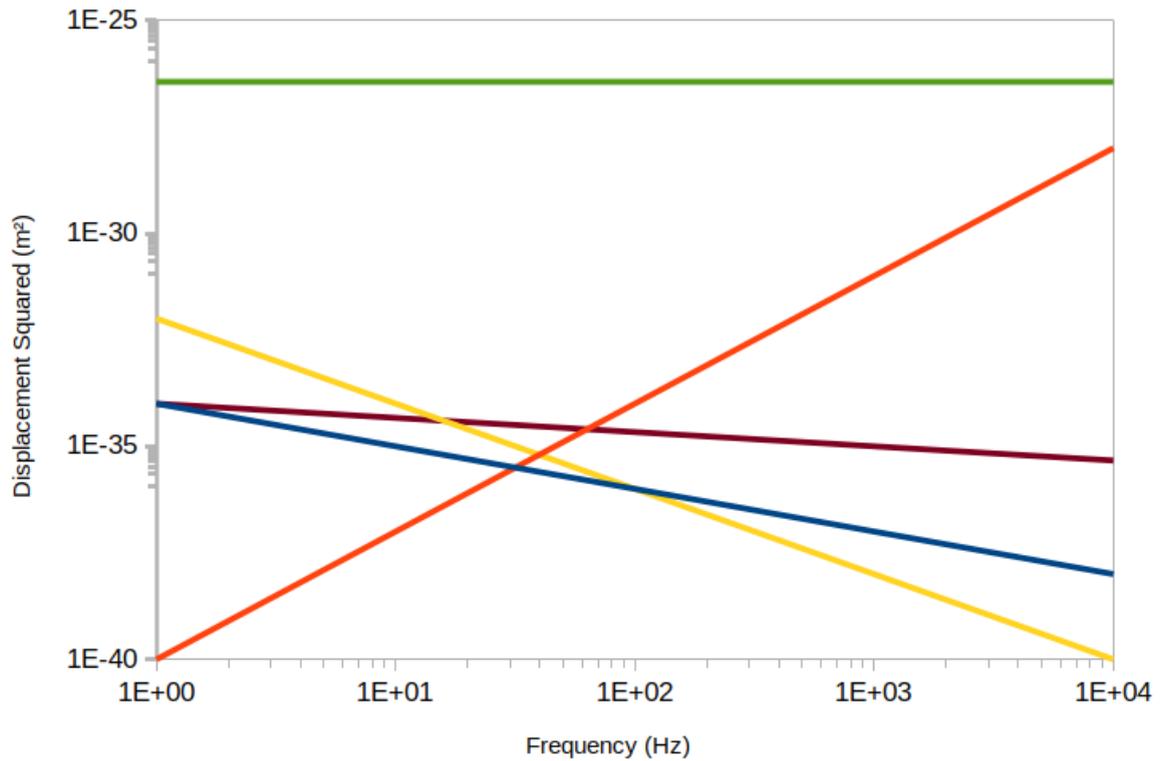

Figure 8. In this figure the amplitude of the dependence of noise, limits and signal can be seen. The green line is the displacement signal in the detector, the red line is the detector sensitivity, the purple line shows the back-action and the electronic series noise chosen to be equal, the yellow line displays the thermal noise and the blue line shows the quantum limit. Source: From the authors.

## Acknowledgments

CF acknowledges CNPq for grant #309098/2017-3 and # FAPESP (Brazil) for grants #2013/26258-4 and #2006/56041-3.